\documentclass[a4paper,12pt]{article}
\usepackage{jheppub2} % for details on the use of the package, please
\usepackage[T1]{fontenc} % if needed

\author[a]{Marco Bochicchio}
\affiliation[a]{INFN sez. Roma 1\\Piazzale A. Moro 2, Roma, I-00185, Italy}

\emailAdd{marco.bochicchio@roma1.infn.it}

\abstract{  
The solution of the large-N 't Hooft limit of QCD is universally believed to be a String Theory of Closed Strings in the Glueball Sector and of Open Strings in the
Meson Sector. Yet, we prove a no-go theorem, that the large-N limit of QCD with massless quarks, or more generally, that the large-N limit of a vast class of confining, i.e. with a Mass Gap in the Glueball Sector, asymptotically-free Gauge Theories coupled to matter fields with no mass scale in perturbation theory cannot be a canonically-defined String Theory of Closed and Open Strings, i.e. admitting Open/Closed Duality. The no-go theorem occurs because Open/Closed Duality, implying that the ultraviolet divergences of annulus diagrams in the Open Sector arise from infrared divergences of tadpoles of massless particles in the Closed Sector, turns out to be incompatible with the existence of the Mass Gap in the Glueball Sector of confining asymptotically-free theories with no mass scale in perturbation theory in which, as for example in QCD, the first coefficient of the beta function for 't Hooft gauge coupling gets $1/N$ corrections due to the matter fields. Moreover, we suggest a way-out to the no-go theorem on the basis of a new non-canonical construction of the String S-matrix for asymptotically-free Gauge Theories such as large-N QCD, involving Topological Strings on Non-Commutative Twistor Space.}

\usepackage{braket}
\usepackage{amsmath}
\usepackage{amssymb}
\usepackage{graphicx}
\usepackage{hyperref}
\usepackage{fancyhdr}
\def\beq{\begin{equation}}
\def\eeq{\end{equation}}
\def\bea{\begin{eqnarray}}
\def\eea{\end{eqnarray}}
\def\bq{\begin{quote}}
\def\eq{\end{quote}}
\DeclareMathOperator{\Tr}{Tr}

\title{Asymptotic Freedom versus Open/Closed Duality in Large-N QCD}
\date{}
\begin{document}
\maketitle

\section{The no-go theorem} \label{1}

Solving the large-N 't Hooft  limit of QCD \cite{H1} is a long-standing problem with outstanding implications for the theory of strong interactions and perhaps for the physics beyond the standard model, if any, in case it arises by a new strongly-interacting sector. It is universally believed that such a solution may be a String Theory \cite{VR} of Closed Strings in the Glueball Sector and of Open Strings in the Meson Sector \cite{H1, Veneziano0}, although it has not been found for more than forty years, despite the advent of the celebrated Gauge/Gravity Duality \cite{Mal} in the framework of String Theory, that we will comment about later. A quantitative evidence in favor of a String Solution is that the perturbative expansion of large-N QCD, once expressed in terms of 't Hooft gauge coupling $g^2= N g^2_{YM}$, implies the following estimates \cite{H1,Veneziano0} for connected correlators of local single-trace gauge-invariant operators $\mathcal{G}_i(x_i)$ and of fermion bilinears $\mathcal{M}_i(x_i)$, both normalized in such a way that the two-point correlators are on the order of $1$:
\bea \label{P}
&&\langle \mathcal{G}_1(x_1)\mathcal{G}_2(x_2)\cdots \mathcal{G}_n(x_n) \rangle_{conn}\sim N^{2-n} \nonumber \\
&&\langle\mathcal{M}_1(x_1)\mathcal{M}_2(x_2)\cdots \mathcal{M}_k(x_k)\rangle_{conn}\sim N^{1-\frac{k}{2}} \nonumber \\
&&\langle \mathcal{G}_1(x_1)\mathcal{G}_2(x_2)\cdots \mathcal{G}_n(x_n)\mathcal{M}_1(x_1)\mathcal{M}_2(x_2)\cdots \mathcal{M}_k(x_k)\rangle_{conn}\sim N^{1-n-\frac{k}{2}} 
\eea
This is exactly the counting \cite{H1,Veneziano0} that we would get in a Theory of (Oriented) Strings with coupling $g_s \sim 1/N$, of Closed Strings in the Glueball Sector, i.e. for a sphere with $n$ punctures, of Open Strings in the Meson Sector, i.e. for a disk with $k$ punctures on the boundary, and of Open/Closed Strings in the Meson/Glueball Sector, i.e. for a disk with $k$ punctures on the boundary and $n$ punctures in the interior. The punctured disk in the Meson Sector arises perturbatively in 't Hooft expansion from planar diagrams whose boundary is a quark loop. This is the Planar Theory, that contains tree amplitudes. Unitarization \cite{H1,Veneziano0} introduces higher-genus amplitudes that correct additively the Planar Theory with a weight $g_s^{-\chi}$, where $\chi$ is the Euler characteristic of the Riemann surface which the amplitude is supported on. Physically, this is the standard picture of confinement, in which the string world-sheet is identified with a chromo-electric flux tube, Mesons are quark-antiquark bound states linked by the chromo-electric flux, and Glueballs are closed rings of flux. Contrary to the universal belief, we prove a no-go theorem, that the large-N limit of QCD with massless quarks, or more generally, that the large-N limit of a vast class of confining, i.e. with a Mass Gap in the Planar Glueball Sector, asymptotically-free Gauge Theories coupled to matter fields with no mass scale in perturbation theory cannot be solved by a canonically-defined String Theory, i.e. admitting Open/Closed Duality. To say it in a nutshell, the no-go theorem occurs because Open/Closed Duality in large-N QCD with massless quarks turns out to be incompatible with Asymptotic Freedom and Mass Gap in the Planar Glueball Sector. The proof is as follows.
Open/Closed Duality occurs in String Theory for example because the annulus diagram (with any number of punctures), i.e. the one-loop diagram in the Open Sector, is topologically the same as the cylinder (with the same punctures), i.e. a tree diagram in the Closed Sector. Moreover, there is a conformal map, in fact a modular transformation, under which they are mapped into each other. Thus, if conformal symmetry, or just the aforementioned modular transformation, on the world-sheet is not anomalous, i.e. the String Theory really exists, the annulus and the cylinder are identical, in the sense that the annulus in the Open Sector can be interpreted as a cylinder in the Closed Sector. In String Theory the annulus, i.e. the one-loop diagram in the Open Sector, being a one-loop diagram can be either ultraviolet finite or ultraviolet divergent. Let us suppose that is logarithmically ultraviolet-divergent (we prove below that this situation would occur in QCD for the annulus with some punctures). If it is, the ultraviolet divergence occurs when the internal boundary shrinks to a point \cite{V}, the annulus degenerating to a punctured disk. Under the modular transformation that maps the annulus into the cylinder one boundary of the cylinder goes farther apart to infinity as the boundary of the annulus shrinks \cite{V}. Therefore, since the modular transformation exchanges the ultraviolet with the infrared and vice-versa, the annulus can diverge in the ultraviolet if and only if the cylinder in the Closed Sector has an infrared divergence when one of its boundaries goes to infinity, corresponding to the actual propagation of massless Closed String states or, more unlikely, to the existence of an accumulation point at zero mass for an infinite number of states, assuming that there is no tachyon in the Closed Sector, i.e. that the theory makes sense. Equivalently, the semi-infinite cylinder being topologically a disk with one puncture in its interior, the states corresponding to the massless (scalar) tadpole occur in the Closed Sector: Indeed, the annulus, the cylinder and the once-punctured disk are suppressed with respect to the disk precisely by a factor of $1/N$, that is the appropriate weight for Closed String states according to Equation \ref{P}. Therefore, if the annulus is ultraviolet divergent in the Open Sector, the String Theory cannot have a Mass Gap in the tree Closed Sector, and vice-versa. All these features have been explicitly verified in different frameworks in a number of papers (see the detailed review \cite{DV}): The modular map that maps the ultraviolet of the annulus into the infrared of the cylinder has been explicitly constructed, and Open/Closed Duality, in particular when the annulus is ultraviolet divergent and the dual infrared divergence occurs in the Closed Sector, has been explicitly checked \cite{DV}. Moreover, it has been suggested that Open/Closed Duality in certain circumstances \cite{DV} is in fact the explanation of Gauge/Gravity Duality: In several cases explicitly worked out the Gauge side is realized by an Open String with a divergent annulus diagram in order to reproduce the correct beta function \cite{DV}, and Open/Closed Duality is used to construct the Gravity Dual. As a result, although the Gravity Dual may be asymptotically free, it is not confining at tree level in these examples, i.e. there is no Mass Gap in the Gravity Sector (i.e. in the tree Closed String Sector) \cite{DV}.
In fact, the relation between the ultraviolet of the annulus and the infrared of the cylinder is a particular case of a much more general correspondence \cite{V}: "The world-sheet of a multi-loop graph of Open Strings is conformally equivalent to a Closed String tree graph. All boundaries in the Open String graph can be represented in the dual channel by external Closed String states, equal to suitable D-brane boundary states. By this dual representation all potential ultraviolet divergences of the Open String graph are equivalent to potential infrared divergences arising from on-shell Closed String States in the dual channel." In particular, we can adapt our statements verbatim to any Planar diagram in QCD with the insertion of a boundary, i.e. of a Meson loop. For example, a boundary on a punctured sphere would correspond in QCD to a Meson loop that renormalizes the Glueball tree amplitudes. We prove below that these diagrams as well would be in general logarithmically ultraviolet-divergent in QCD. But they are conformally equivalent to tree Closed String diagrams. Therefore, since the aforementioned conformal map exchanges the ultraviolet with the infrared \cite{V}, the corresponding tree Closed String diagrams must be logarithmically infrared-divergent because of a massless tadpole. The same conclusion follows, that a Mass Gap cannot occur in the tree Closed String Sector, contradicting the assumption for large-N QCD with massless quarks. We now prove that in the large-N 't Hooft limit of QCD with massless quarks the sum of non-planar diagrams with the topology of the annulus with some punctures, for example in the Flavor-Singlet Meson Sector, and the diagrams obtained adding one Meson loop to the Planar Theory, are in general logarithmically ultraviolet-divergent. Hence the large-N limit of QCD with massless quarks does not admit a canonically-defined String Solution, unless the tree Dual Closed Sector, i.e. the would-be Planar Glueball Sector, does not have a Mass Gap, i.e. it does not confine, contrary to the assumption \footnote{We stress that there is no possibility that the massless states needed for the supposed Open/Closed Duality in large-N QCD with massless quarks be identified with the Goldstone Bosons (i.e. the massless pions) of the spontaneously-broken chiral-flavor symmetry or with their flavor-singlet counterpart (i.e. the $\eta'$, massless to the leading $1/N$ order), which are believed to occur in the Planar Theory under the same assumptions of the no-go theorem: The pions and the $\eta'$ are Mesons and thus occur in the Open rather than in the Closed Sector of the Planar Theory. Besides, they are pseudoscalars.}. Proof: On the Gauge side, since QCD with massless quarks remains exactly massless to all orders of perturbation theory because of chiral symmetry, by dimensional transmutation the only scale is
$\Lambda_{QCD}$, and all the masses of the Glueballs and of the Mesons are pure numbers times $\Lambda_{QCD}$.
Therefore, on the String side the only String parameter is the String Tension $T=\Lambda_{QCD}^2$, where the equality holds up perhaps to a multiplicative constant that is just an inessential change of renormalization scheme. The question arises: Does the String Tension in large-N QCD get ultraviolet-divergent contributions in the $1/N$ expansion ?. The answer is negative in the pure Gauge Theory, i.e. in the pure Glueball Sector, i.e. in the Closed Sector alone. But it is positive for Gauge Fields plus Massless Quarks, i.e. in the Meson Sector, i.e. in the would-be Open String Sector, as follows by expanding $\Lambda_{QCD}$ in powers of $1/N$ and using Asymptotic Freedom. Here it is the computation. 
By definition $\Lambda_{QCD}= const \Lambda \exp(-\frac{1}{2\beta_0 g^2}) (\beta_0 g^2)^{-\frac{\beta_1}{2 \beta_0^2}}(1+...)$ with $\beta_0= \beta_0^P + \beta_0^{NP}= \frac{1}{(4 \pi)^2} \frac{11}{3} - \frac{1}{(4 \pi)^2} \frac{2}{3} \frac{N_f}{N}$ and $\beta_1= \beta_1^P + \beta_1^{NP}= \frac{1}{(4 \pi)^4} \frac{34}{3} - \frac{1}{(4 \pi)^4} (\frac{13}{3} - \frac{1}{N^2}) \frac{N_f}{N}$, where the superscripts $P$ and $NP$ stand for planar and non-planar, and the dots represent terms that vanish as $g \rightarrow 0$. In the Planar Theory, i.e. at the leading $1/N$ order in QCD, and in the Pure Glueball Sector (i.e. for $N_f=0$) to all the $1/N$ orders, the first-two coefficients of the beta function $\beta_0, \beta_1$ get contributions only from planar diagrams. This implies that in large-N QCD without quarks the $1/N$ expansion of the String Tension is in fact finite, the non-planar $1/N$ corrections in the Pure Glueball Sector occurring in the dots contributing only at most a finite change of renormalization scheme to the Planar String Tension $\sqrt {T^P} = \Lambda^P_{QCD} \sim \Lambda \exp(-\frac{1}{2\beta^{P}_0 g^2}) (\beta^{P}_0 g^2)^{-\frac{\beta^{P}_1}{2 \beta_0^{P2}}}(1+...)$. This matches on the QCD side the expectation that Closed String Theories are ultraviolet finite in general because of modular invariance of Closed String diagrams.
But when quark loops are added to on the Gauge side, the first coefficient of the beta function $\beta^P_0$ gets an additive non-planar $1/N$ correction $\beta_0^{NP}$. As a consequence 
$\Lambda_{QCD}$, i.e. the square root of the would-be String Tension $\sqrt{T}$ on the String side, expanded around the Planar Theory gets a non-planar logarithmically-divergent contribution: 
\bea \label{alpha}
\sqrt T &\sim & \Lambda \exp(-\frac{1}{2\beta^{P}_0 (1+\frac{\beta_0^{NP}}{\beta_0^{P}}) g^2 })
\sim  \, \, \Lambda \exp(-\frac{(1-\frac{\beta_0^{NP}}{\beta_0^{P}})}{2\beta^{P}_0  g^2 }) \nonumber \\
&\sim&  \, \, \Lambda \exp(-\frac{1}{2\beta^{P}_0  g^2 })(1+\frac{\frac{\beta_0^{NP}}{\beta_0^{P}}}{2\beta^{P}_0  g^2 }+ \cdots)  \nonumber \\
&\sim& \Lambda^P_{QCD} (1+\frac{\beta_0^{NP}}{\beta_0^{P}} \log(\frac{\Lambda}{\mu})+ \cdots) 
=  \, \, \sqrt{T^P} (1+\frac{\beta_0^{NP}}{\beta_0^{P}} \log(\frac{\Lambda}{\mu})+ \cdots) 
\eea
where in the first line $g$ is a bare free parameter according to the renormalization group of QCD to all the $1/N$ orders, while in the last line we have renormalized $g$ according to the Asymptotic Freedom of the Planar Theory $\frac{1}{2\beta^{P}_0  g^2} \sim \log(\frac{\Lambda}{\mu})$, as follows for consistency by requiring that the Planar Theory has a finite String Tension $T^P$ as $\Lambda \rightarrow \infty$. From 't Hooft perturbative $1/N$ diagrammatic expansion it is clear why $\Lambda^P_{QCD}$ receives non-planar logarithmically-divergent corrections: Were not the case, $\Lambda^P_{QCD}$ would get from the would-be Open Sector only a finite renormalization, implying that the Planar Theory and the theory with one more loop of quarks have the same beta function, that is false in QCD. Indeed, the extra quark loops with respect to the Planar Theory, that come from the quark functional determinant, give origin to a logarithmic divergence due to the non-planar change of the beta function that they introduce. This divergence has to be removed by a further counter-term that is absent in the Planar Theory, since the would-be String Solution and the non-planar 't Hooft expansion must have the same ultraviolet divergences, after the Planar Theory has been renormalized both on the Gauge and on the String side by means of the dimensional transmutation of $g$ into $\Lambda^P_{QCD}=\sqrt{T^P}$.
This is a physical fact, that characterizes how much hard in the ultraviolet Meson loops are, reflecting the corresponding ultraviolet behavior of non-planar quark loops in 't Hooft expansion. 
Hence, being $\sqrt T$ the only mass scale, Meson masses squared receive $1/N$ logarithmically-divergent self-energy corrections proportional to the one of $T$, that can arise only from a logarithmic divergence of the annulus diagrams (with some punctures). The same argument applies to the renormalization of Glueball masses due to a Meson loop. In particular, the large-N limit of QCD with massless quarks may neither be a canonically-defined String Theory, nor a finite theory in the $1/N$ expansion. On the contrary, $T$ would stay finite had we first re-summed all the powers of $N_f/N$ in Veneziano limit of QCD \cite{Veneziano0}, that takes into account already at leading order the non-planar correction to the beta function, but Veneziano Theory, as opposed to the Planar Theory, does not involve tree amplitudes at leading order, and it is therefore presently outside the most remote limits of our methods. This proof of the no-go theorem extends to any confining (supersymmetric or not), i.e. with a Mass Gap in the Planar Closed Sector, asymptotically-free theory with no perturbative mass scale in which the first coefficient of the beta function receives $1/N$ corrections from the would-be Open Sector. Moreover, it does not depend on the details of the Open/Closed String Theory, i.e. on the specific realization of its branes, extra dimensions and so on, but only on the universal features of the conformal theory underlying the String world-sheet. Thus it cannot be evaded in the canonical String framework. Strictly speaking the no-go theorem does not affect the Pure Closed String Sector, and therefore it does not affect Gauge/Gravity Duality in itself. However, though the no-go theorem tells us nothing about the large-N 't Hooft limit of pure Yang-Mills, it does not only to the extent we do not pretend to couple the Closed Sector, i.e. Gravity, to Open Strings too. Yet, in the String framework there is nothing that would forbid it, rising therefore doubts, but it is a matter of opinions, also on the existence of a canonically-defined String Theory solving only the pure Gauge Sector of confining asymptotically-free theories (supersymmetric or not). Indeed, Gauge/Gravity is in fact presently totally empty for confining asymptotically-free gauge theories, because there is actually no such a model based on Gauge/Gravity Duality that is Asymptotically-Free and has a Mass Gap, i.e. that confines and reproduces in the Closed String Sector the asymptotics of the two-point connected correlator of the action density in an asymptotically-free theory, as extensively discussed in \cite{MBM,MBN}:
\bea
\int\sum_{\alpha\beta\gamma\delta}}{\langle  \Tr{F_{\alpha\beta}^2}(x) \Tr{F}_{\gamma\delta}^2(0)\rangle_{conn}e^{ip\cdot x}d^4x  &\sim& \frac{p^4}{\beta_0\log\frac{p^2}{ \Lambda_{QCD}^2}}\Biggl(1-\frac{\beta_1}{\beta_0^2}\frac{\log\log\frac{p^2}{  \Lambda_{QCD}^2 }}{\log\frac{p^2}{ \Lambda_{QCD}^2  }}\Biggr)  \nonumber \\
&\sim& \sum_{k=1}^{\infty} \frac{  g^4(m_k^2) m_k^4  \rho^{-1}(m^{2}_k)  }{p^2+m^{2}_k}
\eea
where $\rho(m^2)$ is the unknown scalar spectral density. Summarizing, it is impossible to construct a canonically-defined String Theory that solves the large-N 't Hooft limit of QCD with massless quarks. The easy way-out to the no-go theorem is to give up Asymptotic Freedom, and to declare that the (canonical) String is an effective description only in the infrared. For example, there exists Luscher \cite{L1} computation of the universality class of Wilson loops in the infrared, and its most recent refinement \cite{L2}, based once more on Open/Closed Duality \cite{L2}. However, guided by Asymptotic Freedom \cite{MBH}, we suggested a new non-canonical construction of the generating functional of the one-loop S-matrix $\Gamma$, based on Topological Strings on Non-Commutative Twistor Space \cite{MBH, MBL}:
\bea
\Gamma &=&-\sum_p \Tr \log(1-P \exp i \int_{C_p} \hat B_{\lambda}) =- \sum_p \hat \Tr \log(\frac{d}{d\lambda}+ \hat B_{\lambda})|_{C_p} \nonumber \\
&=&-\log Det(\frac{d}{d\lambda}+ \hat B_{\lambda})
\eea
obtained re-summing the expansion in powers of Wilson loops, that arises by world-sheet instantons supported on Riemann surfaces of fixed genus, into a functional determinant \cite{MBH} expanded around $\hat B_{\lambda}= \tilde B_{\lambda} + g_s \delta \tilde B_{\lambda}$, that therefore is not supported on Riemann surfaces of fixed genus for any fixed number of external states, thus evading the no-go theorem, which we were unaware of in \cite{MBH}, in a number of ways.

\section{Acknowledgments}

We would like to thank Gabriele Veneziano for enlightening explanations on Planar Duality, that was the initial framework of this work, and for several discussions about the no-go theorem.
We would like to thank Massimo Bianchi for clarifying discussions on the no-go theorem. We would like to thank Adi Armoni as well, for a deep discussion on the no-go theorem after this paper was posted in the arXiv, and for pointing out to us that somehow similar arguments existed for vacuum diagrams \cite{S,N,A}, relating a tachyon in the infrared of the annulus to an exponential density of states in the ultraviolet of the cylinder \cite{A}.

\end{document}